# Superluminal Correlations in Ensembles of Optical Phase Singularities

T. Bucher[1†], A. Gorlach[1†], A. Niedermayr[1], Q. Yan[1], H. Nahari[1], K. Wang[2], R. Ruimy[1], Y. Adiv[1], M. Yannai[1], T. L. Abudi[1], E. Janzen[3], C. Spaegele[4], C. Roques-Carmes[5], J. H. Edgar[3], F. H. L. Koppens[6,7], G. M. Vanacore[8], H. H. Sheinfux[9], S. Tsesses[1,10], I. Kaminer[1*]

[1]*Andrea & Erna Viterbi Department of Electrical and Computer Engineering, Technion–Israel Institute of Technology, 3200003 Haifa, Israel.*
[2]*Shanghai Institute of Optics and Fine Mechanics, Chinese Academy of Sciences, Shanghai 201800, China.*
[3]*Tim Taylor Department of Chemical Engineering, Kansas State University, Manhattan, KS 66506, USA.*
[4]*Harvard John A. Paulson School of Engineering and Applied Sciences, Harvard University, Cambridge, MA, USA.*
[5]*E. L. Ginzton Laboratory, Stanford University, 348 Via Pueblo, Stanford, CA, 94305, USA*
[6]*ICFO-Institut de Ciències Fotòniques, The Barcelona Inst. of Sci. and Tech., Av. Carl Friedrich Gauss 3, 08860 Castelldefels, Spain.*
[7]*ICREA-Institució Catalana de Recerca i Estudis Avancats, Passeig Lluís Companys 23, 08010 Barcelona, Spain.*
[8]*Department of Materials Science, University of Milano-Bicocca, Milano 20126, Italy.*
[9]*Department of Physics, Bar-Ilan University, Ramat Gan 5290002, Israel*
[10]*Department of Physics, MIT-Harvard Center for Ultracold Atoms and Research Laboratory of Electronics, Massachusetts Institute of Technology, Cambridge, MA, USA*

[†]Equal contribution
*Corresponding author: kaminer@technion.ac.il

**Phase singularities—points carrying quantized topological charge—are universal features found across diverse wave systems from superfluids and superconductors to acoustic and optical fields[1-5]. Ensembles of such singularities exhibit distance correlations resembling particles in liquids[6-9], extensively studied for their role in exotic material phases[10-12]. In contrast, the full correlations in phase-space that govern the system evolution have remained unexplored and experimentally inaccessible. Here, we directly measure the ultrafast dynamics of optical singularity ensembles, capturing their full phase-space correlations, presenting the joint distance-velocity distribution. Our observations reveal a breakdown of the particle-singularity analogy[13]: phase singularities exhibit acceleration to unbounded velocities before annihilation[8,14-17], indicated by measurements of velocities exceeding the speed of light. Intriguingly, these superluminal velocities are paradoxically amplified by the slow group velocity of hyperbolic phonon polaritons in our material platform, hexagonal boron nitride membranes[18-22]. We demonstrate these phenomena using combined hardware and algorithmic advances in ultrafast electron microscopy[23-31], achieving spatial and temporal resolutions each an order of magnitude below the polaritonic wavelength and cycle period. Our findings enable probing topological defect dynamics at previously unattainable timescales, deepening our understanding of phase-singularity universality and suggesting phenomena of ultrafast information flow in polaritonic media.**

## Introduction

Singularities of many kinds arise in nearly every branch of physics, from dislocations in crystals[10] and flux quanta in superconductors[1], to vortex cores in fluid flows (e.g., cyclones)[2] and quantized vortices in superfluids[3]. Other manifestations are ubiquitous in materials supporting optical, phononic, and plasmonic wave fields[4]. The understanding of singularities has roots in the 1885 "hairy ball theorem"[32], and has vastly evolved ever since, particularly in the context of singular optics[5,33,34]. Optical singularities enable precise control of light-matter interactions with both bound[35-37] and free[38-40] electrons, underpin super-resolution imaging[41-44], and serve as carriers of classical[45] and quantum information[46]. These opportunities motivate extensive research into generating and imaging singularities, continually uncovering their fundamental properties across a wide range of optical systems[12,47-51].

Another reason for the interest in singularities in wave systems is the strong analogy between phase singularities and interacting particles. The particle-singularity analogy arises due to the stable nature of phase singularities carrying topological charge $\pm 1$, characterized by a $\pm 2\pi$ phase winding. Rare singularity events could, in principle, carry higher integer topological charges, but they are typically unstable[52,53]. Much like particle-antiparticle pairs, a singularity annihilates only when encountering another singularity of the opposite charge. This analogy is regarded as an intriguing manifestation of wave-particle duality in classical systems[6,13,33,54,55].

When dealing with many such singularities, their collective correlations become crucial for understanding the global system properties[11,56,57]. The distance correlations resemble those of interacting particles that make up liquids, independently of the true underlying physics, as shown in pioneering experiments in microwave[58] and optical[7] domains.

Previous studies focused on the distance-correlation functions in singularity ensembles, following Berry's foundational work on their statistical properties[6,59] and later works on singularity-pair correlations[8]. Extensive theoretical efforts have been dedicated to analyzing the dynamics of singularity ensembles under temporal evolution[8,14-17,34], including the prediction of their velocity distributions[8]. However, experimental research has lagged behind theory. Observing the dynamical statistical properties requires measuring the optical (or quantum) system with a sub-cycle, sub-wavelength resolution, which is a significant technical challenge.

Remarkable phenomena in optical singularity dynamics remain hidden, both at the level of individual singularities and within their collective ensembles[33]. Most notably, theory has long predicted that optical singularities can exhibit superluminal motion, particularly at moments close to their creation or annihilation, where their velocities can become unbounded[8,14-17,34].

Here, we monitor the ultrafast dynamics of optical phase singularities with deep sub-wavelength spatial and deep sub-cycle temporal resolutions, revealing their acceleration near annihilation events. Quantitative analysis of the singularity velocities reveals transient superluminal motion. We explore the singularities' ultrafast dynamics by both direct observation of individual annihilation events and by using large-scale statistical analysis in the phase-space of their position and velocity, captured by distance-velocity correlations among all singularity pairs. The measured distance correlations support the well-known particle-like nature of singularities, whereas their velocity distribution pinpoints the breakdown of the particle analogy via superluminal speeds and sub-cycle creation and annihilation events.

**Deep Sub-Wavelength and Deep Sub-Cycle Mapping of Phase Singularities**

The key findings emerge from reaching deep sub-wavelength phase imaging (20 nm ~ $\lambda_{\text{PhP}}/30$), with deep sub-cycle temporal resolution (3 fs ~ $T/8$), over a large field-of-view and long overall measurement duration. This experiment is performed on an optical platform of hyperbolic phonon-polariton (PhP) wavepackets in a thin hexagonal boron nitride (hBN) flake[18,6058], using an ultrafast transmission electron microscope (UTEM)[23]. We rely on the photon-induced near-field electron microscopy (PINEM) technique[23-27], extended to resolve the field phase[28-31] via free-electron Ramsey imaging (FERI)[29], providing unambiguous identification of the singularity charge. We employ mid-infrared femtosecond pulses (average wavelength $\lambda_0 = 7$ μm, corresponding to a field cycle of $T = 23.3$ fs) to create phonon polariton (PhP) wavepackets. Because of the hyperbolic dispersion characteristic of PhPs[18], they can be confined to sub-wavelength dimensions, in our experiments on the order of $\lambda_{\text{PhP}} \approx \lambda_0/11 \approx 630$ nm. The hyperbolic dispersion also leads to a slow group velocity, reaching well over 100 times slower than the speed of light[19-21]. Furthermore, PhPs exhibit remarkably low optical losses, especially for isotopically pure hBN, for which the PhP lifetime can reach well over a picosecond[22]. The combination of all these factors, with the advance of phase-resolved polariton imaging in electron microscopy, makes hBN a favorable platform to observe universal properties of singularity dynamics.

Each iteration of the experiment begins by dividing the near-infrared pulse into three paths (Fig. 1(a)). The first path is upconverted to the ultraviolet regime (via fourth harmonic generation) and used to emit the probe electrons in the UTEM. The rest of the pulse is down-converted (via difference frequency generation) to the mid-infrared range, corresponding to hBN's upper Reststrahlen band supporting PhPs. The second path is used to pre-modulate the electrons at a reference sample. The third path generates the PhP wavepackets at the hBN sample. Details of the experimental scheme appear in the supplementary materials (SM), Section I.

The reference and sample pulses can be tuned independently, allowing control over their relative sub-cycle (phase) delay $\Delta\varphi$, intensity, and polarization. By measuring multiple sub-cycle delays $\Delta\varphi$, and the algorithm from 29, we extract the phase and amplitude (Fig. 1(b,c)) of the PhP near-field component pointing along the electron trajectory. To image also the group dynamics of the PhP, we change the time delay $\Delta t$ between the electron pulses and the light excitation of the sample. For each time delay, we reconstruct the amplitude and phase, enabling us to track the field's group dynamics (Fig. 1(d) and SM movie).

`

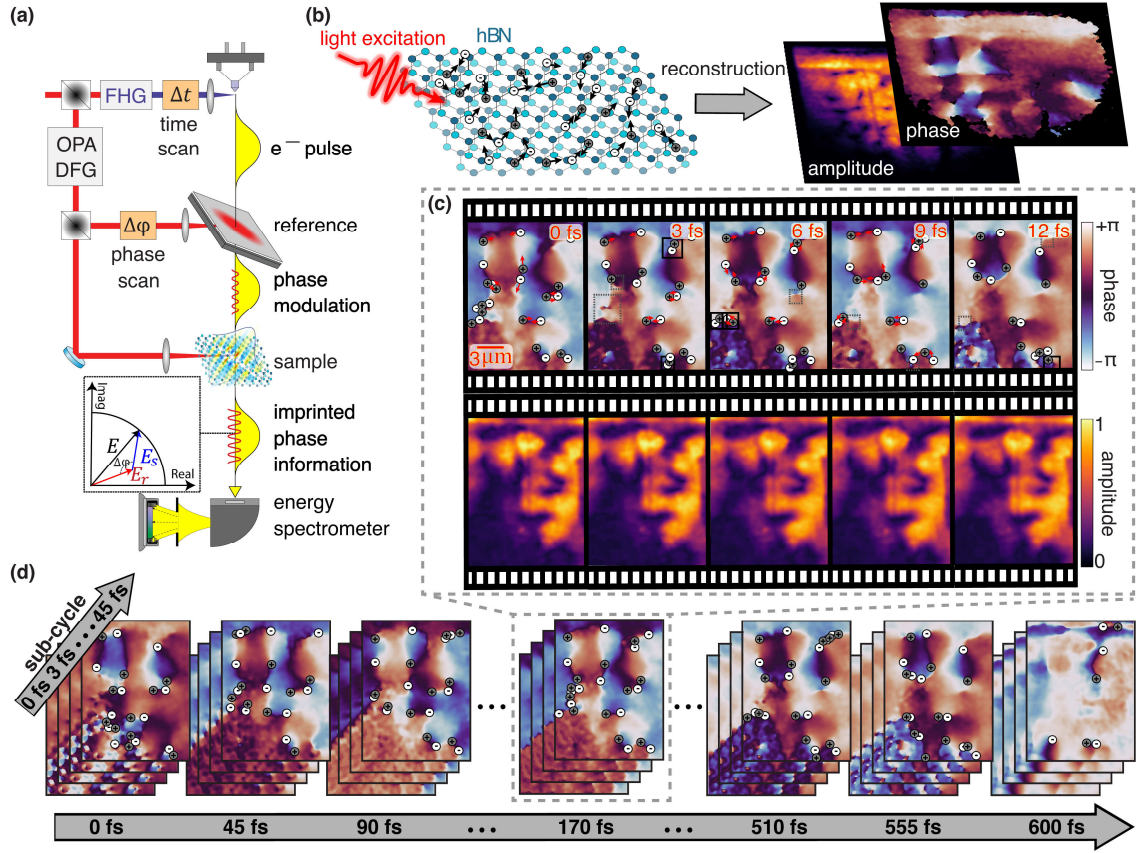

**Fig. 1. Deep sub-wavelength deep sub-cycle imaging of optical phase singularities in hexagonal boron nitride (hBN), recording both phase and group dynamics.** (a) A femtosecond laser pulse is split into three parts to excite the electron pulse, modulate it at the reference, and create PhPs in the sample. By timing the pulses, we record the phase ($\Delta\varphi$) and group dynamics ($\Delta t$). (b) The spatiotemporal dynamics of singularities and their phase-space correlations are measured in hBN, extracting their phase and amplitude. (c) Measured phase dynamics of singularities show sub-cycle creation and annihilation events. (d) The entire measurement (see also SM movie) captures both deep sub-cycle (3 fs) and group dynamics (>800 fs) of singularities, with deep sub-wavelength resolution (20 nm) over a macroscopic field of view (21×21 µm²). These constitute deep sub-wavelength phase imaging (×350 below the free-space wavelength and roughly ×30 below the PhP wavelength), with deep sub-cycle temporal resolution (×8 below the cycle time).

Each retrieved frame contains a complex interference pattern, involving phase singularities of right- and left-handed $2\pi$ winding, corresponding to positive or negative topological charge[8]. After aligning all the frames through computational reconstruction, we automate the identification of singularities, separating them by positive and negative topological charges (SM Section V, Fig. S2). Analyzing this extensive dataset, we also automate the tracking of the singularities' trajectories, denoting their creation and annihilation events. By quantifying singularity positions and velocities over time, we analyze their statistical properties, specifically deriving their velocity distribution and joint distance-velocity distributions, constituting the full phase-space correlations.

A common framework for studying numerous properties of phase singularities and their correlations is the interference of Gaussian random 2D waves[6,7,8,52,61]. Despite its simplicity, this model captures a wide range of universal wave phenomena and applies to diverse experimental settings. Indeed, as we see below by comparing the theory with our experimental results, this model successfully describes many advanced features of PhP wavepackets in hBN. Moreover, the model holds even when we extend the theoretical analysis to predict the joint distance-velocity correlations, offering a complete characterization of singularity dynamics within their phase space.

**Deep Sub-Cycle Annihilation of Singularity Pairs with Superluminal Velocities**

The acceleration of opposite-charged singularities before annihilation or after creation is a universal feature in the interference of Gaussian random waves and can be understood through the space-time trajectories of annihilating singularities (Fig. 2(a)). As opposite-charged singularities approach each other, their paths in space-time must form a continuous curve at the annihilation point, forcing their acceleration to unbounded velocities right before the annihilation. This fact is a mathematical consequence of the continuity of the phase rather than a violation of physical laws: Phase singularities carry zero intensity and thus can "move" superluminally without energy (or information) transmission[13].

The speed with which singularities move must be distinguished from other ubiquitous superluminal phenomena in optics[62], such as the well-known superluminal propagation of optical phase fronts[63] or the superluminal Rabi rotation of optical vortex cores[64,65]. Unlike the phase-front dynamics, for phase singularities, the unbounded velocities correspond to their universal space-time trajectories as they accelerate toward annihilation.

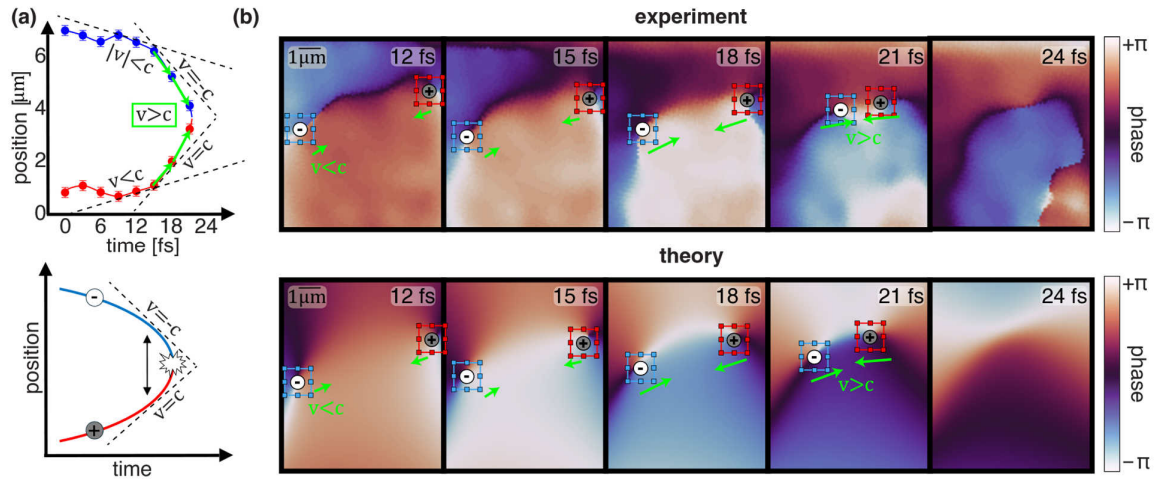

**Fig. 2. Deep sub-cycle annihilation of singularities, showing an example of acceleration to unbounded velocities along a characteristic space-time trajectory. (a)** The trajectories of oppositely charged singularities approaching annihilation form a closed, continuous curve in space-time, shown in experimental results (top) and illustration (bottom). Near annihilation, any such continuous trajectory must approximate a parabolic curve, which necessarily contains

a region where the velocity of the singularity pair is unbounded and consequently exceeds the speed of light (denoted by a dashed tangent). **(b)** (top) Measurement of singularity dynamics within 9 fs, focusing on the final frames before annihilation, in which the singularity dynamics is superluminal. (bottom) Simulation based on the Gaussian random wave model, showing a qualitatively similar annihilation event.

This acceleration of the singularities before annihilation is not unique to optical singularities and can, in principle, be observed in various physical systems. In superfluids, vortex-antivortex pairs accelerate toward each other before annihilating, with their velocity increasing sharply just before the collision[66]. In superconductors, magnetic vortices and antivortices experience mutual attraction and accelerate rapidly before annihilation, producing a characteristic voltage peak[67]. Similarly, in fluid dynamics, vortex rings can accelerate and deform as they approach collision, exhibiting a burst of speed before merging or breaking apart[68]. These examples stress that the characteristics of phase singularities are universal across different physical platforms, in optics and beyond. While similar pre-annihilation acceleration has been observed in superfluids, superconductors, and fluid vortices[66-68], in all these platforms, the velocities remained subluminal.

Below, a large-scale statistical analysis of the measured singularities helps identify what conditions made the optical PhP platform favorable for the observation of superluminal dynamics. We monitor the entire sample area (21 × 21 µm²) over 800 ps, analyzed in 285 phase-resolved frames (each created by 15 sub-cycle frames). An ensemble of ~50 singularities is tracked in each frame. This large dataset allows us to quantify *dynamical statistical properties and correlations* among all singularity pairs, revealing universal collective properties of singularity ensembles.

**Distance and Velocity Correlations within Ensembles of Singularities**

We compare simulations of the Gaussian random wave model with our experimental results by deriving distance and velocity correlation functions for singularities (Fig. 3(a,b,c)). Fig. 3(d) presents the experimentally measured and theoretically predicted distance correlation functions $g_{+|+}(R) = g_{-|-}(R)$ and $g_{+|-}(R) = g_{-|+}(R)$, defined as the probability of finding a pair of singularities at a distance $R$ from one another, for the same or opposite charge[7,8,61]. The in-plane isotropy of hBN PhPs implies that the distance correlations depend only on the distance $R$ with no dependence on the polar angle. The measured distance correlations $g_{+|+}$ and $g_{+|-}$ match the theoretical prediction of the Gaussian random wave model, with larger error bars at smaller distances due to the scarcity of events with small distances.

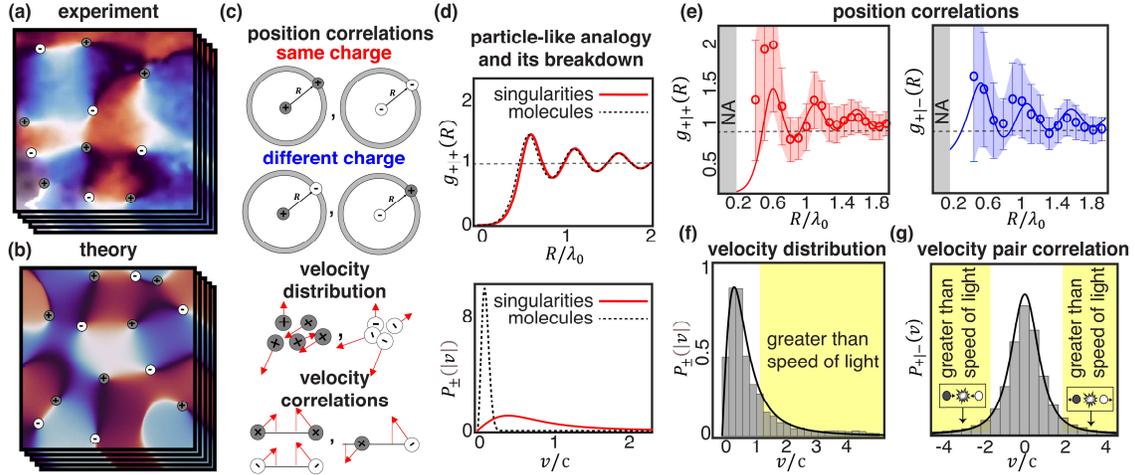

**Fig. 3. Distance correlations and velocity distributions of singularities. (a)** Experimentally measured phase of the PhP field in an hBN flake, displaying "+" and "-" singularities whose phase wraps by ±2π. **(b)** Similar singularities appear in the phase of a simulated interfering Gaussian random waves. **(c)** Definition of the distance correlation (top) for same-charge and different-charge pairs, and of velocity distribution and correlations (bottom). **(d)** Theoretically predicted (solid lines) and experimentally measured (circles) distance correlation functions $g_{+|+}(R)$ ($g_{+|-}(R)$). The experimental data has larger error bars for small $R$ because there are fewer events when singularities are at close distances to each other (See SM, Section VI). N/A denotes regions that do not contain enough singularities to generate trustworthy statistics. **(e)** The distance correlation (top) of singularities resembles that of interacting particles that make up liquids[9], whereas the singularity velocities distribution (bottom) breaks this analogy, showing a large fraction of singularities being superluminal. Hence, while singularities share some particle-like properties, they remain a unique system. **(f)** Theoretical and experimental singularity velocity distributions $P_{\pm}(|v|)$. The average velocity of the singularities is $\langle v \rangle = 3.12 \times 10^8$ m/s $\approx 1.04c$, where $c$ is the speed of light in vacuum. Part of the singularities have velocities much higher than the speed of light. **(g)** Theoretical and experimental relative velocity correlations show a good match. Unbounded negative velocities correspond to annihilation events, whereas unbounded positive velocities correspond to creation events.

Fig. 3(e) presents an analysis of the ubiquitous particle analogy for singularities: As also shown before[7], the distance correlation function resembles those of particles that make up a liquid (with average period of correlations ~0.3 nm [9]), exhibiting similar spatial short-range order due to their interactions (Fig. 3(d)). Going beyond the distance correlations, we now also measure the velocity distribution (Fig. 3(f)), as predicted in 8. The velocity distribution breaks the particle-singularity analogy (Fig. 3(e)), exactly as discussed in 8. Specifically, the distribution emphasizes the unique behavior of singularities ensembles, exhibiting a much longer tail compared to the (Maxwell–Jüttner) velocity distribution of particles at the same relativistic temperatures (Fig. 3(e) bottom).

The velocity distribution of phase singularities $P_{\pm}(|v|)$ is defined as the probability density of singularities to have absolute velocity $|v|$ (with $P_+(|v|) = P_-(|v|)$ due to symmetry):

$$P_{\pm}(|v|) = \int_0^{2\pi} P_{\pm}(v_x = |v|\cos\theta, v_y = |v|\sin\theta)|v|d\theta. \quad (3)$$

Furthermore, the velocity distribution can be calculated analytically[8], showing good agreement

with the experimental results in Fig. 3(f):

$$P_{\pm}(|v|) = \frac{8\pi^2 \langle v \rangle^2 |v|}{(\pi^2|v|^2 + 4\langle v \rangle^2)^2}. \tag{4}$$

$\langle v \rangle$ is the average velocity of the singularities and is measured directly in our experiment to be $\langle v \rangle \approx (1.04 \pm 0.004)c$, in close agreement with the theoretical prediction[8]: The average velocity is given by $\langle v \rangle = c(1 + (k_{PhP}/\Delta k)^2)^{-1/2} \pi/\sqrt{2}$, where the standard deviation in wavenumber $\Delta k$ satisfies $k_{PhP}/\Delta k = \lambda_0/\left((v_{ph}/v_g)\Delta\lambda\right)$, with $v_{ph}$ and $v_g$ being the average phase and group velocities of the PhPs (see SM, Section X). In hBN, the group velocity of PhPs is much smaller than the phase velocity over a broad-spectral range, with $v_{ph}/v_g \approx 12 \pm 1$ in our case, leading to a theoretical prediction of $\langle v \rangle \approx (1 \pm 0.1)c$, matching our measurement.

This result highlights a unique property of hBN PhPs: their slow group velocity allows the average singularity velocity to approach or even slightly exceed the speed of light. In comparison, the average velocity would be much lower if not for the hyperbolic nature of hBN PhP. For instance, in free space, where $v_{ph} = v_g$, the average singularity velocity is an order of magnitude smaller $\langle v \rangle \sim 0.1c$. Consequently, only 0.4% of the singularities would exceed the speed of light, for the same laser parameters (analysis shown in Fig. S4). In contrast, 29% of the singularities in our system exceed the speed of light, as shown by both data and theory in Fig. 3(f). This comparison stresses that the example we provided in Fig. 2 is not exceptional, but in fact very common in our experiment. Thus, the PhP platform makes it far more likely to observe superluminal events.

**Full Phase-Space Correlations within Ensembles of Singularities**

We next extend the theory of singularities in Gaussian random waves. We first predict the characteristic velocity correlations among singularity-pairs, for same-charge $P_{+|+}(v)$ and opposite-charge $P_{+|-}(v)$ singularities. Analytical analysis is presented in SM, Sections VII-IX. These correlations represent the probability of two singularities having a relative velocity $v$ along their connecting line.

$$P_{\sigma|\sigma'}(v) = \frac{1}{N_{\sigma\sigma'}} \langle \sum_{a\in\sigma, b\in\sigma'} \delta(v - (\boldsymbol{v}_a - \boldsymbol{v}_b) \cdot \boldsymbol{R}_{ab}) \rangle, \tag{5}$$

where $\sigma, \sigma' \in \{+, -\}$ specify charge types, $N_{\sigma\sigma'}$ is the total number of pairs with such charges, $\widehat{R}_{ab}$ is a unit vector along the line connecting charge $a$ with $b$, and $\delta$ is the Dirac delta function. The brackets $\langle ... \rangle$ denote an average over all possible realizations. This definition leaves us with two distinct probabilities: same charge $P_{+|+} = P_{-|-}$ or opposite charges $P_{+|-} = P_{-|+}$.

Fig. 3(g) presents $P_{+|-}(v)$, demonstrating good agreement with theory. The correlations become charge-independent ($P_{+|+}(v) = P_{+|-}(v)$) for large-enough sample sizes (shown in SM, Section VII). i.e., analysis of the entire singularity ensemble remains unaffected by creation and annihilation events, which are rare. To capture the full dynamics of the singularities and especially pinpoint their behavior close to annihilation and creation events, we next consider a more advanced correlation function.

Finally, we define the full phase-space correlations, described by the joint distance-velocity distribution that captures the entire system's evolution:

$$P_{\sigma|\sigma'}(v,R) = \frac{1}{N_{\sigma\sigma'}(R)} \langle \sum_{a\in\sigma, b\in\sigma'} \delta(v - (\boldsymbol{v}_a - \boldsymbol{v}_b)\cdot \boldsymbol{R}_{ab}) \delta(R - |\boldsymbol{r}_a - \boldsymbol{r}_b|) \rangle, \quad (6)$$

$N_{\sigma\sigma'}(R)$ is the total number of pairs with charges $\sigma\sigma'$ separated by distance $R$. Symmetry considerations imply that $P_{+|+}(v,R) = P_{-|-}(v,R)$ and $P_{+|-}(v,R) = P_{-|+}(v,R)$.

Fig. 4 presents both theoretical predictions (analytical results in SM, Section IX) and experimental results obtained by analysis of the entire dataset, showing good agreement. The joint probabilities $P_{+|+}(v,R)$ and $P_{+|-}(v,R)$ generalize the analysis in previous works[8,16]: The distance correlations are reproduced once integrating over $v$, and the velocity distributions are reproduced once integrating over $R$.

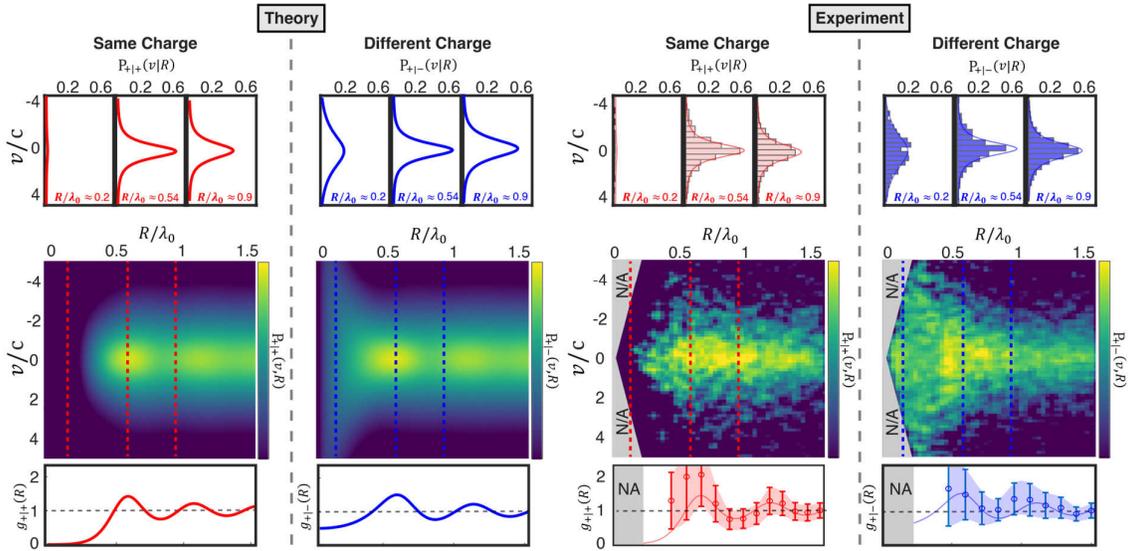

**Fig. 4. Full phase-space correlations of singularities.** Plots of $P_{+|+}(v,R)$ and $P_{+|-}(v,R)$ as a function of both $v$ and $R$ show a good match between theory and experiment, revealing distinct behaviors: At proximity, same-charge singularities ($P_{+|+}(v,R)$) are uncommon. In contrast, for the same $R$ values, opposite-charge singularities ($P_{+|-}(v,R)$) are more common and exhibit higher velocities, indicating acceleration before annihilation or after creation. Experimental results (right) align well with the theoretical model (left) of random wave interference. N/A denotes regions where not enough singularities were found to generate trustworthy statistics or where the velocities were too high for detection.

Fig. 4 shows that at small distances ($R < \lambda_0$), the fraction of created and annihilated singularity pairs is large, leading to higher overall values of $P_{+|-}(v,R)$ and a wider variance, indicating higher possible velocities. The complementary $P_{+|+}(v,R)$ is smaller, since same-charge singularities are less likely at small distances, as expected by the known instability of singularities with charges higher than "±1" [52,53]. At larger distances, singularities cannot be created or annihilated, resulting in narrower velocity distributions (also see SM, Fig. S3). Finally, from comparison to theory, it is evident that the maximum observable velocity is limited not by the material or polariton field, but by the state-of-the-art of microscopy and the maximum spatial and temporal resolution.

**Conclusion and Outlook**

We experimentally observed dynamical correlations among ensembles of optical singularities in hBN, using ultrafast electron microscopy. Our measurements captured deep sub-cycle (3 fs $\approx T/8$) creation and annihilation events of singularities with deep sub-wavelength resolution (20 nm $\approx \lambda_{\mathrm{PhP}}/30$). This combined resolution in time and space, in conjunction with the long interrogation time and large field of view, enabled us to demonstrate long-standing predictions about singularity velocity correlations and the possibility of unbounded velocities[8], a universal phenomenon for Gaussian random waves. The imaging capabilities presented in this work motivated more advanced statistical analysis of dynamical correlation functions. Specifically, measuring the joint distance-velocity distributions revealed the acceleration of oppositely charged singularities before annihilation and after creation.

The past decade has shown renewed interest in the superoscillatory[42-44,69,70] nature of certain wavefields—where local field/phase gradients can exceed the maximum spatial frequency. Superoscillations play a role near the center of every phase singularity[71], with prospects for advanced optical microscopy techniques with deep subwavelength resolution[41]. In our context, the unbounded velocities of singularities we observe are a direct manifestation of superoscillatory field gradients, once undergoing temporal evolution. Conversely, the rapid creation and annihilation of singularities inherently generate superoscillations, highlighting a deep connection between the extreme dynamics of topological defects and extreme superoscillatory features of the wavefield.

The experimental approach presented here could extend to the study of polaritons in other 2D materials and their heterostructures, which would be particularly interesting in systems exhibiting highly tunable optical properties. Polaritonic properties of these materials include exotic topological phases and intricate coupling of the electric and magnetic fields[72,73], all suggesting intriguing changes in the singularity distributions and correlations. Materials with strong nonlinear optical responses could break the Gaussian random wave model and enable the generation of more complex interference patterns, potentially containing novel types of singularities beyond the fundamental ±1 orders. Higher-order and multi-dimensional singularities provide a larger and richer space for encoding information[74,75], which could also correspond to more intricate types of collective phenomena in the overall system.

Free electrons can not only probe the dynamics of singularities, but their wavefunctions can also be modified through interactions with these singularities. Such interactions can imprint phase singularities onto electron wavefunctions, generating novel electronic states[38-40,76], with potential applications in electron holography and other electron interference techniques.

Finally, the analytical approaches developed here could be leveraged to address long-standing challenges in electron microscopy, such as the fluctuating granularity or "bee-swarm" effect[77,78]. By quantitatively mapping the dynamics of these fluctuations and analyzing their ensemble-level correlations, we can devise mitigation strategies that push atomic-scale imaging beyond current limits.

## Methods

### Ultrafast transmission electron microscope (UTEM)

Experiments were performed in a pump–probe UTEM based on a JEOL JEM-2100 Plus with a LaB6 electron gun operated at 200 kV (Extended Data Fig. 1a-b). The microscope was run in low-magnification mode (objective lens off) to keep the electron beam paraxial (convergence angle < 1 mrad). Post-column electron energy loss spectrometer (EELS) (0.1 eV dispersion) provided energy filtering; a slit selected electrons that gained energy in both interactions. Images were recorded on a direct-detection camera (Gatan K2 Summit). The zero-loss peak full-width half maximum (FWHM) in vacuum was ~1.4 eV.

### Laser system, optical paths, and timing

A 40 W, 1030 nm, ~270 fs, 1 MHz laser (Carbide) drove the pump–probe scheme. One branch was frequency-converted to 266 nm to photo-emit single-electron probe pulses at the LaB6 cathode. A second branch was difference-frequency-converted to the mid-IR (~7 µm) and split into two: one illuminated the sample and the other the reference interaction. Two delay stages controlled (i) the pump–probe delay for group-dynamics mapping and (ii) a fine sub-cycle phase delay between reference and sample interactions. Mid-IR pulses were TM-polarized and focused to ~100 µm (sample; 4–12 mW average power) and ~500 µm (reference; 4–20 mW). The sample path entered via a side port and the sample was tilted by ~35° to avoid shadowing. The reference path impinged above the sample at ~20° with the reference membrane tilted by ~41° to satisfy phase matching with ~0.7 c electrons.

### Photonic electron–light modulator (PELM) and energy-filtered imaging

The reference interaction used a modified Hard X-ray Aperture module carrying $Si_3N_4$ membranes (Extended Data Fig. 1c) coated with ~25 nm Al to implement a pre-sample photonic electron-light modulator (PELM)[79]. Electrons interacted first with the reference field and then with the sample field before energy selection and imaging. The average absorbed power at the sample was below reported heating/damage thresholds for hBN[80].

### Free-electron Ramsey imaging (FERI) and phase retrieval

We used free-electron Ramsey imaging (FERI)[29] to reconstruct the complex near field at the sample. By scanning the relative optical phase between reference and sample interactions in sub-cycle steps and recording energy-filtered images, we retrieved amplitude and phase per pixel via an optimization-based forward model of the PINEM/FERI interaction.

### Data acquisition, reconstruction, and drift correction

We recorded 285 raw frames over ~855 fs with 3 fs sub-cycle sampling. Each amplitude/phase reconstruction used 15 sequential raw measurements; adjacent reconstructions overlapped by 14 frames to preserve temporal resolution. To correct mechanical and beam drift, we performed semi-automated feature extraction and applied affine transforms frame-by-frame before subsequent analysis (Extended Data Fig. 2). A linear interpolation produced a 0.2 fs-step movie for downstream singularity tracking.

**Sample preparation**

Isotopically pure h11BN crystals were grown as detailed in 81, then mechanically exfoliated using low-retention PDMS (dry transfer). Flakes were transferred onto 20 nm SiN membranes (Norcada). Thicknesses of 40–50 nm were confirmed by EELS log-ratio analysis. The hBN flake covered most of the $21 \times 21$ $\mu m^2$ field of view; sharp flake edges acted as near-field couplers for launching hyperbolic phonon-polaritons (PhPs).

**Singularity identification, charge assignment, and tracking**

From the reconstructed phase maps, we located optical phase singularities by evaluating the $2\pi$ phase winding around pixel loops and assigned topological charge $\pm 1$. Connected-component clustering yielded singularity centroids for positive and negative charges. We enforced periodic boundary conditions when computing pair distances. Inter-frame association used a linear-assignment (Hungarian) approach with distance thresholds to obtain trajectories, creation, and annihilation events; velocities were computed by finite differences along tracks.

**Distance correlations, velocity distributions, and phase-space analysis**

We computed same-charge and opposite-charge distance-correlation functions $g_{+|+}(r) = g_{-|-}(r)$ and $g_{+|-}(r) = g_{-|+}(r)$ as probabilities of finding singularity pairs at separation $r$. [8] Velocity statistics included the singularity speed distribution $P(v)$ and relative-velocity correlations along the inter-singularity axis. To capture full dynamics, we evaluated the joint distance–velocity distribution $P(r, v)$, which marginalizes to $g(r)$ and $P(v)$. Experimental observables were compared with a theory of isotropic Gaussian random waves (non-monochromatic extension), for which we derived two-point correlators including temporal derivatives and obtained $P(r, v)$ via Gaussian sampling of the joint distribution.

*Further derivations, and validations are provided in the Supplementary Information (Materials and Methods, Sections I–X).*

**Statistics and reproducibility**

Distance and velocity distributions were aggregated over all reconstructions; uncertainties reflect finite-sample statistics, with larger error bars at small separations due to the scarcity of near-collision events. Regions without sufficient counts are marked N/A in figures.

**Estimating characteristic velocities and role of group velocity**

The average velocity equals to:

$$\langle v \rangle = c \frac{\pi}{\sqrt{2}} \frac{\Delta k/k}{\sqrt{1 + (\Delta k/k)^2}},$$

where $k$ is the average wavenumber of phonon-polaritons (PhPs) in the hBN sample, $\Delta k$ is the standard deviation of phonon-polaritons. In the experiment, the average wavelength and its standard deviation are $\lambda_0 = 7$ μm and $\Delta \lambda = 0.3$ μm respectively.

The standard deviation in $k$ is connected with the standard deviation in wavelength according to $\Delta k/k \approx (v_{\text{ph}}\Delta\lambda)/(v_{\text{g}}\lambda_0)$. Thus, the $\Delta k/k$ is much larger than the initial deviation in wavelength $\Delta\lambda/\lambda_0$ due to the unique property of PhPs in the sample to have a group velocity $v_{\text{g}} \ll v_{\text{ph}}$. In our case, we have $v_{\text{ph}}/v_{\text{g}} \approx 12 \pm 1$. This leads to the following average velocity:

$$\langle v \rangle \approx (1 \pm 0.1)\, c,$$

which is in very good agreement with the experimental result $\langle v \rangle \approx (1.04 \pm 0.004)c$. We want to emphasize that we achieved superluminal velocities in the experiment due to the unique property of PhPs to have very slow group velocities. If the group velocity were equal to the phase velocity, then we would have $\langle v \rangle \approx 0.1c$. However, due to $v_{\text{ph}}/v_{\text{g}} \approx 12 \pm 1$, we get an average velocity higher than the speed of light $c$.

**Image processing and software**

All detector-level processing was disabled; raw camera data were exported and processed with in-house MATLAB code implementing the FERI forward model, phase retrieval, drift correction, singularity detection/tracking, and correlation analyses. Pseudocode for Algorithms 1–3 (singularity detection, clustering, and tracking/velocity extraction) is provided in the Supplementary Information.

**Acknowledgments**

This project is funded by the European Union's ERC COG, QinPINEM, Project Number 101125662. We acknowledge funding from the Helen Diller Quantum Center. This research is also funded by the Gordon and Betty Moore Foundation, through Grant GBMF11473. This work is part of the SMART-electron project, which has received funding from the European Union's Horizon 2020 Research and Innovation Programme under Grant Agreement No 964591. S.T. acknowledges generous support from the Adams fellowship of the Israeli Academy of Science and Humanities; the Yad Hanadiv foundation through the Rothschild fellowship; the VATAT-Quantum fellowship by the Israel Council for Higher Education; the Helen Diller Quantum Center post-doctoral fellowship; and the Technion Viterbi fellowship. E.J. and J.H.E were supported by the Office of Naval Research, Award N00014-20-1-2474. CRC was funded by a Stanford Science Fellowship. K.W. was supported by the National Natural Science Foundation of China (No. 12374321), the Shanghai Rising-Star Program (No. 22QA1410100), and the international partnership of Chinese Academy of Sciences (111GJHZ2022024FN).

# Extended data figures and tables

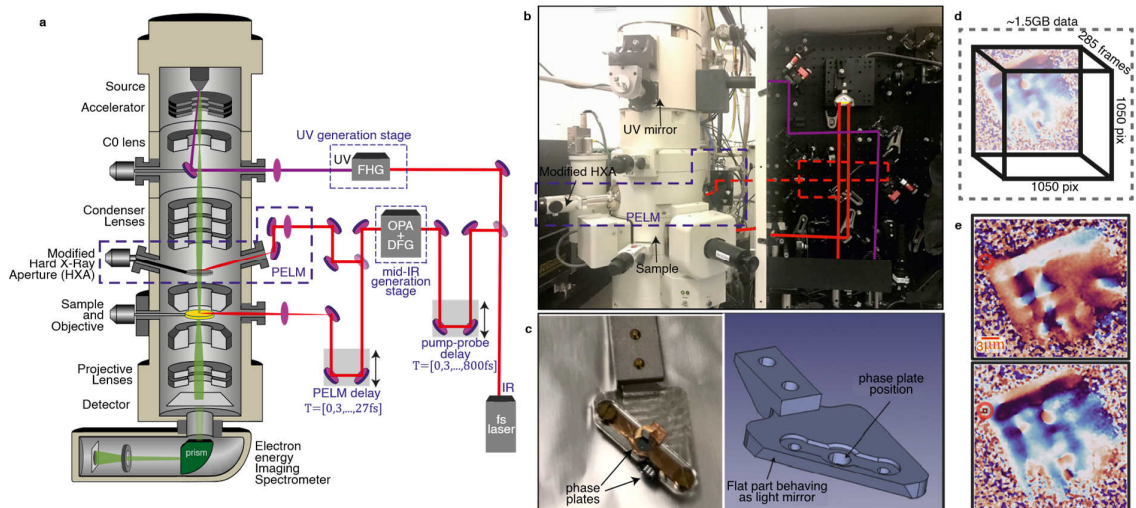

**Extended Data Fig. 1 | The UTEM setup and the PELM integration.** UTEM illustration **(a)** and image **(b)** illustrating the microscope column, electron spectrometer and detectors, optical setup, and the integration of a modified Hard X-ray Aperture (HXA) at a post-condenser lens stage (PELM). The external knob of the HXA (a and b, left side) has two rigid positioning points with 5 mm lateral travel around them for positioning the reference interaction point with respect to the electron beam path. An electron-transparent thin film sits at the place of the x-ray aperture, and light enters from the optical access port on the opposite side of the column at a 20-degree angle above the horizon (dashed red line in b). Double illumination scheme (a and b, right side) implemented on the vertical board next to the UTEM. The IR laser beam is separated into two portions using a 50:50 beam splitter. One portion is guided towards the PELM (dashed red line in b), whereas the other portion is guided towards the sample (solid red line in b). **(c)** Image (left) and CAD model (right) of the modified HXA aperture connected to the platelet hosting the electron-transparent light-opaque metallic thin films for electron-light interaction. The platelet is made of Aluminum alloy, whereas the clamp is made of 0.15-mm-thick Beryllium Copper. One can observe two Si-window TEM grids (Norcada Inc.), which are coated with a 25-nm-thick Aluminum film deposited via thermal evaporation on a 10-nm-thick $Si_3N_4$ membrane. In each grid, nine slots are present to maximize the available points of interaction in case of local damage to one of the membranes. The platelet has also been cut at a specific angle, allowing it to host a small metallic mirror able to reflect the light down the column towards the sample position (not used in the current work). The platelet, HXA, and their integration were designed and performed in close collaboration with IDES, part of JEOL Ltd. **(d)** By using the pump-probe delay stage in combination with the PELM delay stage, the setup allows a very long acquisition time in high spatio temporal resultion with a large field of view. The result is 285 frames of 1050x1050 pix images, a total size of ~1.5GB of data to analyze with our specialized algorithmic process. **(e)** The very long acquisition time also results in sample and beam instability, which needs to be taken into account.

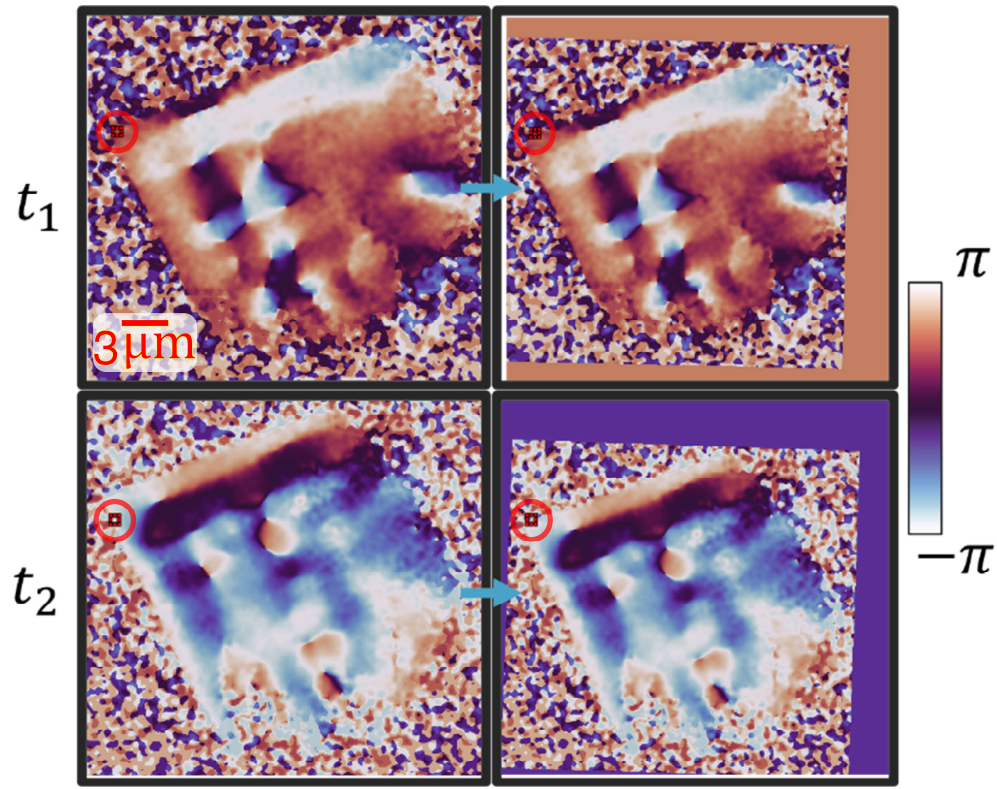

**Extended Data Fig. 2 | Correcting sample and beam drift.** Left: phase reconstructions for two different times $(t_1, t_2)$, each time has a different rotation and translation, which is fixed by calculating an affine transformation from at least 10 features selected manually on each frame. Right: corresponding fixed phase reconstruction. The red circle marks the same pixel indices, which points on different coordinates on the sample for different times before the correction (left). After correction, the rectangle marks the same coordinates on the sample.